\documentclass[
reprint,
superscriptaddress,
aps,
prd
]{revtex4}
\usepackage[english]{babel}

\usepackage[letterpaper,top=2cm,bottom=2cm,left=3cm,right=3cm,marginparwidth=1.75cm]{geometry}
\usepackage{subfigure}
\usepackage{amsmath}
\usepackage{graphicx}
\usepackage[colorlinks=true, allcolors=blue]{hyperref}
\begin{document}

\title{ Baryogenesis and CMB spectral
distortion  from Axions}

\author{Zhenhao Zhang}
\email[]{zhangzhenghao6830@163.com}
\author{Mingqiu Li}
\email[]{limingqiu17@mails.ucas.ac.cn}
\affiliation{School of Physics, Beijing Institute of Technology, Beijing, 100081, China}
\author{Sichun Sun }
\email[]{sichunssun@gmail.com}
\affiliation{School of Physics, Beijing Institute of Technology, Beijing, 100081, China}
\begin{abstract}
We discuss a mechanism for generating the baryon asymmetry in the early universe. We show that an axion-like particle can modify the related gauge field configurations in the Standard Model, thereby altering their dispersion relations. This change in the Chern–Simons number can source a violation of baryon number. We derive the relationship between the resulting baryon number and the evolution of the axion background. We estimate the baryon asymmetry produced via this mechanism and show that the observed value can be naturally achieved.  We also show that axion–photon coupling produces Cosmic Microwave Background spectral distortion. Our results show that the resulting distortion approaches a constant at low frequencies, unlike the conventional y-type and $\mu$-type distortions.

\end{abstract}

\maketitle
\section{ Introduction}	
The origin of baryon asymmetry is a longstanding mystery. Research on both the anisotropy of the cosmic microwave background and big bang nucleosynthesis indicates that the baryon number density normalized by the entropy density is  $n_B/s\approx 9\times 10^{-11}$ \cite{Planck:2018nkj, ParticleDataGroup:2024cfk}.  This asymmetry can be produced in the early universe through various theoretical mechanisms, but none have been confirmed by experiments.  It is commonly believed that any asymmetries are diluted by inflation; also, a large initial baryon asymmetry before inflation predicts correlated baryon isocurvature perturbations that are already excluded by cosmic microwave background observations \cite{Murai:2023ntj}.   The baryon asymmetry should normally be produced after inflation, with some exceptions \cite{Kusenko:2014uta}.
 %It is commonly believed that any asymmetries are diluted by inflation; therefore, the baryon asymmetry should be produced after inflation. 
 %Besides,  a large initial baryon asymmetry before inflation predicts correlated baryon isocurvature perturbations that are already excluded by cosmic microwave background observations \cite{Murai:2023ntj}.  
 To explain the baryon asymmetry, many mechanisms have been proposed, including electroweak baryogenesis \cite{Morrissey:2012db}, baryogenesis via leptogenesis \cite{Strumia:2006qk}, baryogenesis from decaying magnetic helicity \cite{Kamada:2016eeb}, axion-induced baryogenesis \cite{Jeong:2024hhi,Co:2019wyp, Kusenko:2014uta}, etc. In ref \cite{Kamada:2016eeb}, the evolution of a primordial hypermagnetic field interacting with turbulent plasma leads to decaying helicity and finally transfers to baryons. In this work, we discuss a mechanism by which the axion-like particle(ALP) can change the helicity of the gauge field in the Standard Model (SM) and thus can generate baryon asymmetry. Our model is initially similar to the axion-induced model \cite{Jeong:2024hhi,Co:2019wyp,Kusenko:2014uta}, but differs in the mechanism for generating baryon number.

The axion was introduced to solve the strong CP problem \cite{PhysRevLett.38.1440, PhysRevD.16.1791}. It has been generalized into a broader concept that addresses many open questions, such as dark matter \cite{Jiang:2021dby, Adams:2022pbo}, inflation \cite{Pajer:2013fsa}, and the propagation of cosmological gamma rays \cite{DeAngelis:2007dqd}. ALP is a Nambu-Goldstone boson of a general $U(1)_{\rm PQ}$ symmetry breaking which can couple to $SU(2)_L\times U(1)_Y$ gauge fields in the Standard Model (SM) with Chen-Simons terms. Because a non-vanishing velocity of a classical axion field spontaneously breaks CPT \cite{Domcke:2020kcp,Schmitz:2015nqa}, baryon asymmetry can be generated in thermal equilibrium without fulfilling Sakharov’s conditions \cite{Sakharov:1967dj}. Axions can lead to an energy shift between particle and antiparticle \cite{Shi:2015zwa}, and thus, if there exist baryon number-violating interactions, baryon asymmetry can be generated \cite{Jeong:2024hhi,Co:2019wyp,Kusenko:2014uta}. In ref \cite{Kusenko:2014uta}, heavy right-handed neutrinos were induced to violate lepton number and finally lead to baryogenesis by sphalerons during inflation. In ref \cite{Jeong:2018jqe, Jeong:2024hhi}, the ALP   enables a strong first-order phase transition due to the coupling with the Higgs and thus can trigger electroweak baryogenesis. In ref \cite{Co:2019wyp}, PQ asymmetry is converted to baryon asymmetry by sphaleron transitions, assuming the quarks have $U(1)_{\rm PQ}$ charge.  ALPs can convert to a vector field through narrow resonance \cite{Kitajima:2023pby} and tachyonic instability \cite{Machado:2018nqk}. It can also explain the cosmic birefringence \cite{Luo:2023cxo, Nakagawa:2025ejs, Gasparotto:2023psh,Finelli:2008jv} through splitting the dispersion relation of two circularly polarized photons In this work, unlike the previously complicated models, we focus on the interaction between ALPs and the $U(1)_Y$ gauge field. we directly calculate the baryon number using anomaly equation $\partial_{\mu}J_{B}^{\mu}=\frac{-N_{F}}{32\pi^{2}}g_{Y}^{2}Y^{\mu\nu}\widetilde{Y}_{\mu\nu}$, where ${Y^{\mu\nu}}$ is field strength of $U(1)_Y$ symmetry.  We assume that the gauge fields are always in thermal equilibrium in the early universe, and that the ALPs affect them by modifying the dispersion relation. we derive the relation between $\langle Y^{\mu\nu}\widetilde{Y}_{\mu\nu}\rangle$ and the axion background. The baryon asymmetry due to anomaly is estimated, and we find that the traditional misalignment mechanism \cite{Visinelli:2009zm} can not obtain the observed $n_B/s$ due to the small velocity of the axion field, while the kinetic misalignment mechanism \cite{Co:2019jts} can achieve the desired baryon asymmetry.

Cosmic Microwave Background (CMB) spectral distortions are tiny deviations of the CMB frequency spectrum from a perfect blackbody.
Conventionally, they arise from energy injection or extraction processes in the early universe. Two types of spectral distortions have been extensively studied: the $\mu$ type distortion and the $y$ type distortion \cite{Chluba:2025wxp, Lucca:2019rxf,Chluba:2018cww}. At redshifts $2\times 10^5\lesssim z \lesssim 2\times 10^6$ \cite{Khatri:2012tw}, the double Compton scattering and bremsstrahlung are inefficient at driving the chemical potential of photons to zero, while the Compton scattering is still efficient at redistributing the energy. Any energy injection or extraction will lead to a nonzero chemical potential of photons, or equivalently, a $\ mu$-type spectral distortion. At redshifts $z\lesssim 1.5\times 10^4$ \cite{Khatri:2012tw}, the Compton scattering is too weak to redistribute the energy of photons, and thus y distortion can happen in this epoch. In this work, we show that the axion background can produce CMB spectral distortions. Due to the Chen-Simons coupling between ALPs and photons, the axion background affects the frequency distribution of photons by altering the dispersion relation. We estimate the spectral distortions induced by ALPs and find that their shape differs from that of the conventional y-type and $ \mu$-type distortions. Our result provides new insights for the study of the Chern–Simons couplings to photons.

The paper is organized as follows. In section \ref{sec2} we discuss the gauge fields in an axion background and derive the relation between $\langle Y^{\mu\nu}\widetilde{Y}_{\mu\nu}\rangle$ and the axion background. The baryogenesis is studied in section \ref{sec3}. 
In section \ref{sec5}, we study the CMB spectral distortion from ALPs. 
Section \ref{sec4} is the conclusions and discussions. Throughout the article, we use natural units, i.e., $\hbar = c = k_B = 1$.

\section{gauge fields in axion background}\label{sec2}
Here, we discuss the gauge field value in the presence of the axion modification. An ALP $\phi$ can couple to gauge particles in the SM by Chen-Simons terms. The Lagrangian is given by
\begin{equation}\label{couple}
 L\supset- \beta_1\frac{\phi}{4 f_{a}} Y_{\mu\nu}\tilde{Y}^{\mu\nu}- \beta_2\frac{\phi}{4 f_{a}}W^{i\mu\nu}\widetilde{W}_{\mu\nu}^{i},  
\end{equation}
Where $Y_{\mu\nu},W^i_{\mu\nu}$ are field strength for $U(1)_Y$ and $SU(2)_L$ symmetries in SM respectively. In this work, we focus on the $U(1)_Y$ gauge field; thus, we chose $\beta_2=0$. In an expanding universe,  the  equation of motion for $U(1)_Y$ gauge field  is \cite{Kitajima:2023pby}
\begin{equation}
\ddot{\boldsymbol{Y}}+H\dot{\boldsymbol{ Y}}-\frac{\nabla^{2}\boldsymbol{Y}}{a^{2}}-\frac{\beta_1}{f_{a}a}\left(\dot{\phi}\nabla\times \boldsymbol{Y}-\nabla\phi\times(\dot{\boldsymbol{Y}}-\nabla Y_{0})\right)=0,    
\end{equation}
where $a$ is the scale factor, $H=\dot{a}/a$ is the Hubble parameter, with the overdot indicating derivative with respect to cosmic time. The gauge field can be decomposed into Fourier modes
\begin{equation}
 \boldsymbol{Y}(t,x)= \sum_{\lambda=\pm}\int\frac{d^{3}k}{(2\pi)^{3}}\boldsymbol{e}_{\lambda}(k)e^{i \boldsymbol{k}\cdot \boldsymbol{x}}Y_{\lambda}(t,k), 
\end{equation}
where $\boldsymbol{e}_{\pm}$ are circular polarization vectors satisfying $ \boldsymbol{k}\cdot \boldsymbol{e}_{\pm}=0, \boldsymbol{k}\times \boldsymbol{e}_{\pm}=\mp ik\boldsymbol{e}_{\pm}. $ Assuming a homogeneous axion background, the  equation of motion for Fourier modes is
\begin{equation}\label{eqYY}   \ddot{Y}_{\pm}+H\dot{Y}_{\pm}+\left(\frac{k^{2}}{a^{2}}\mp\frac{k}{a}\frac{\beta_1\dot{\phi}}{f_{a}}\right)Y_{\pm}=0.
\end{equation}
For Sub-horizon modes, i.e.$H\ll k/a$, the dispersion relation is
\begin{equation}\label{ssgx}
  \begin{aligned}
 \omega_\pm&=\left(\frac{k^{2}}{a^{2}}\mp\frac{k}{a}\frac{\beta_1\dot{\phi}}{f_{a}}\right)^{1/2}\\
 &\approx \frac{k}{a}\mp  \frac{\beta_1\dot{\phi}}{2f_{a}}.
  \end{aligned}  
\end{equation}
We have assumed $\beta_1\dot{\phi}/f_a\ll k/a$. The difference between two circular modes can lead to cosmic birefringence \cite{Luo:2023cxo, Nakagawa:2025ejs, Gasparotto:2023psh}. The approximate plane wave solution for Eq. (\ref{eqYY}) is $Y_{\pm}\approx\sqrt{f_{\pm}/k}e^{-i\omega_{\pm}t}$, where $f_\pm$ is a distribution function. 
The energy density for the gauge field is
\begin{equation}
    \begin{aligned}
 \rho_{Y}^{(\pm)}&=\frac{1}{2}\int\frac{k^2d k}{2\pi^2}\left[\frac{|\dot{Y}_{\pm}|^{2}}{a^2}+\frac{k^{2}|Y_{\pm}|^{2}}{a^{4}}\right],\\    &\approx  \int\frac{k^2d k}{2\pi^2} \frac{\omega_\pm|f_{\pm}|}{a^{4}}.
    \end{aligned}
\end{equation}
If the gauge field is in thermal equilibrium, $f_{\pm}$ should take the Bose-Einstein Distribution. Notice that we use a different approach from \cite{Jeong:2024hhi,Co:2019wyp,Kusenko:2014uta}. With this assumption, we have,
\begin{equation}\label{eqfe}
\begin{aligned}
f_{\pm}(k)&=\frac{1}{e^{\omega_\pm/T}-1}\\
    &\approx\frac{1}{e^{k/(aT)}-1} \left(1\pm \frac{\beta_1 \dot{\phi}}{2f_aT}\frac{e^{k/(aT)}}{e^{k/(aT)}-1}\right).\\
\end{aligned}
\end{equation}
Another gauge-independent pseudoscalar is
\begin{equation}\label{edotB}
 \begin{aligned}
  \frac{-1}{4}\langle Y_{\mu\nu}\tilde{Y}^{\mu\nu}\rangle&=\langle\boldsymbol{E\cdot B}\rangle \\
  &=-\frac{1}{a^{3}}\int\frac{d^{3}k}{(2\pi)^{3}}\frac{k}{2}\frac{d}{d t}\left(|Y_{+}|^{2}-|Y_{-}|^{2}\right)\\
  &=-\frac{1}{a^{3}}\int\frac{d^{3}k}{(2\pi)^{3}}\frac{\exp\frac{k}{aT}}{\left(\exp\frac{k}{aT}-1\right)^2}\frac{d}{d t}\left(\frac{\beta \dot{\phi}}{2f_aT}\right)\\
  &=-\frac{\beta_1 T^3 }{12f_a}\frac{d}{dt}\left(\frac{\dot{\phi}}{T}\right).
 \end{aligned}   
\end{equation}
Here, the angular bracket denotes an average over the volume. The analogous electric and magnetic fields are defined by $\boldsymbol{E}=-\dot{\boldsymbol{Y}}/a, \boldsymbol{B}=\nabla\times \boldsymbol{Y}/a^2$ under the temporal gauge $Y_0=0$ respectively. The pseudoscalar $\langle\boldsymbol{E\cdot B}\rangle=-\dot{h}/2$ and $h=\langle\boldsymbol{Y\cdot B}\rangle$ is the magnetic helicity density \cite{Kamada:2016eeb}.  In the next section, we will use this gauge field to generate baryon number.

\section{Baryogenesis}\label{sec3}
In the SM, the $B+L$ anomaly can lead to baryon number violation.  
The  anomaly equation is \cite{Trodden:1998ym}
\begin{equation}
   \partial_{\mu}J_{B}^{\mu}=\frac{N_{F}}{32\pi^{2}}\left(g_{L}^{2}W^{a\mu\nu}\widetilde{W}_{\mu\nu}^{a}-g_{Y}^{2}Y^{\mu\nu}\widetilde{Y}_{\mu\nu}\right), 
\end{equation}
where $N_F=3$ is the number of fermionic generations.
Consider only  $U(1)_Y$ gauge field, the baryon number evolve  as
\begin{equation}
\frac{dn_B}{dt}+3Hn_B=-\frac{1}{V}\int d^3x\frac{N_{F}g_{Y}^{2}}{32\pi^{2}}Y^{\mu\nu}\widetilde{Y}_{\mu\nu}=\frac{N_{F}g_{Y}^{2}}{8\pi^{2}}\langle{\bf E\cdot B}\rangle.
\end{equation}
The gauge fields are always in thermal equilibrium; thus, using Eq.  (\ref{edotB}), the baryon number is related to the axion field by
\begin{equation}\label{baryonN}
 a^3(t_f)n_B(t_f)-a^3(t_i)n_B(t_i)=\frac{ \beta_1  N_f g^2_Y(aT)^3}{96 \pi ^2 f_a}\left(\frac{\dot{\phi _i}}{T_i}-\frac{\dot{\phi _f}}{T_f}\right).    
\end{equation}
Thus, the change of $\dot{\phi}/T$ can lead to baryogenesis. 

The equation of motion for the axion background is
\begin{equation}\label{eqphi}
\ddot{\phi}+3H{\dot{\phi}}+\frac{dV(\phi)}{d \phi}=\frac{\beta}{f_{a}}\langle{\bf E\cdot B}\rangle.
\end{equation}
Without loss of generality, we adopt the following potential
\begin{equation}
    V(\phi)=m^{2}f_a^{2}\left(1-\cos\frac{\phi}{f_a}\right).
\end{equation}
Depending on the axion kinetic energy, the evolution of $\phi$ can be classified into two scenarios \cite{Adams:2022pbo}: the traditional misalignment mechanism and the kinetic misalignment mechanism. In the traditional misalignment mechanism \cite{Visinelli:2009zm}, $\phi$ obtains a non-zero value after inflation, and then rolls to the minima of the potential as the universe expands. Finally $\phi$ oscillates around the minimum. In the kinetic misalignment mechanism \cite{Co:2019jts},  the axion kinetic energy $\dot{\phi}^2/2$ is larger than the maximum potential energy at the conventional oscillation temperature; thus, the axion simply overcomes the potential barrier and continues to change at the rate $\dot{\phi}$. 
\begin{itemize}
\item  traditional misalignment mechanism\\
The equation of motion Eq. (\ref{eqphi}) can be numerically solve with initial conditions $\phi(t_i)=f_a\theta_i,\dot{\phi}(t_i)=0$. 
Using Eq. (\ref{baryonN}), the baryon number normalized by entropy density is
\begin{equation}\label{nBs1}
 \begin{aligned}
  \frac{n_B(t_f)}{s(t_f)}&=-\frac{15 \beta_1  g_Y^2 N_F\dot{\phi}(t_f)}{64 \pi ^4 g_* f_a T_f}\\
  &\simeq -7.8\times 10^{-6}\beta_1\frac{\dot{\phi}(t_f)}{f_a T_f}
 \end{aligned}   
\end{equation}
where the entropy density $s=2 \pi ^2 g_*T^3/45$ with $g_*=106.75 $ is the effective degrees of freedom in SM. We have assumed $n_B(t_i)=0$. Baryogenesis occurs only before the electroweak phase transition. After the electroweak phase transition, the symmetry breaks to $U(1)_{em}$; therefore, the baryogenesis mechanism based on varying helicity of the $U(1)_Y$ gauge field is invalid.

 In this work, we take the critical temperature for the electroweak phase transition $T_c=159.5{\rm GeV}$ \cite{DOnofrio:2015gop} and thus $T_f\geq 159.5{\rm GeV}$.  
Consider a radiation dominated universe, assuming  the initial  misalignment $\theta_i$ is not too large when the initial time $mt_i\rightarrow 0$, Eq. (\ref{eqphi}) has approximate solution
\begin{equation}
\phi(t)=f_a\theta_i2^{1/4}\Gamma\left(\frac{5}{4}\right)(mt)^{-1/4}J_{\frac{1}{4}}\left(mt\right),    
\end{equation}
where $J_{\frac{1}{4}}(x)$ is Bessel function. 

Figs. \ref{fig:two_images} show the $\dot{\phi}/(f_aT)$ as temperature varying with $\theta_i=1$. The oscillation temperature $T_*$ is evaluated by $3H(T_*)=m$, and $\phi$ oscillates when the temperature is lower than $T_*$. In left panel, $T_*\le T_c$, and thus $|\dot{\phi}|/(f_aT)$ increase as temperature decrease before electroweak phase transition. The relation $T_*\le T_c$ gives $m\le 1.1\times 10^{-4}{\rm eV}$. Thus, for $m\le 1.1\times 10^{-4}{\rm eV}$, $|\dot{\phi}|/(f_aT)$ increases as the universe cools before the electroweak phase transition and continually generates baryons. However, as Eq. (\ref{nBs1}) and the specific value in the left panel of Figs. \ref{fig:two_images} indicated, it is difficult to obtain $n_B/s\sim 10^{-10}$ unless we choose a unreasonable large $\beta_1$. In the right panel of Figs. \ref{fig:two_images},  the oscillation temperature $T_*>T_c$ and thus the axion oscillate before  electroweak phase transition. During the electroweak phase transition, $3H\ll m$ means the Compton wavelength of the axion is much smaller than the Hubble scale, and thus the axion field can not be regarded as a homogeneous background. Therefore, for $m\gg 1.1\times 10^{-4}{\rm eV}$, $\dot{\phi}$ at temperature $T_c$ depends on the space positions, and the average generated baryons are zero. In summary, the traditional misalignment mechanism cannot achieve $n_B/s\sim 10^{-10}$. 
%Substituting the solution into Eq. (\ref{nBs1}), The baryon asymmetry $n_B/s$ evolve as temperature decrease is showed in Figs.\ref{fig:two_images}
\begin{figure}[htbp]
    \centering
    \includegraphics[width=0.45\linewidth]{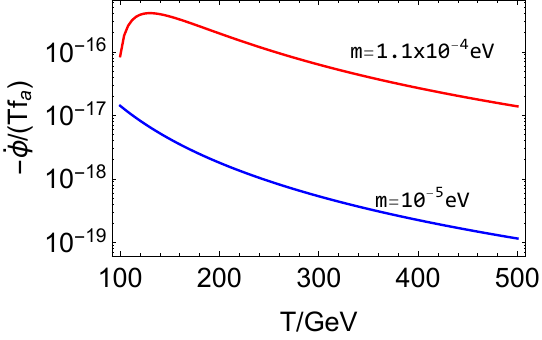}
    \includegraphics[width=0.5\linewidth]{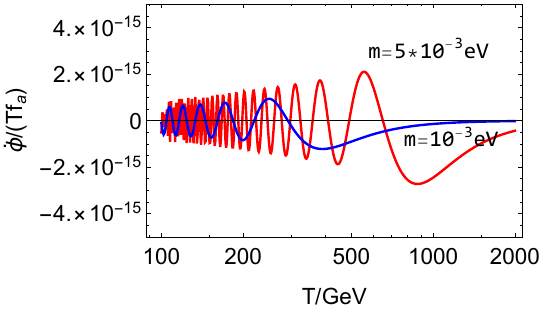}
    \caption{The $\dot{\phi}/(f_aT)$ as temperature varying. In left panel, the oscillation temperature $T_*\le 159.5{\rm GeV}$ while in right panel $T_*> 159.5{\rm GeV}$. We have taken $\theta_i=1$.} \label{fig:two_images} 
\end{figure}

\item kinetic misalignment mechanism\\
In the kinetic misalignment mechanism, the Noether charge associated with the PQ symmetry is $n_{PQ}=\dot{\phi}f_a$. Defining $Y_{PQ}=n_{PQ}/s$, then $Y_{PQ}$ remains constant as the universe expands until the axion kinetic energy is smaller than the potential barrier. The axion abundance nowadays can be estimated as \cite{Co:2019jts}
\begin{equation}\label{eqomega}
\Omega_{a}h^{2}\simeq\Omega_{\mathrm{DM}}h^{2}\left(\frac{10^{9}~\mathrm{GeV}}{f_{a}}\right)\left(\frac{Y_{PQ}}{40}\right),   
\end{equation}
where $\Omega_{\rm DM}= 0.265$ is the dark matter abundance. Using Eq. (\ref{baryonN}), the baryon asymmetry at time $t_f$ is 
\begin{equation}
\frac{n_B(t_f)}{s(t_f)}=\frac{ \beta_1  N_f g^2_YY_{PQ}}{96 \pi ^2 f_a^2}(T^2_i-T_f^2).    
\end{equation}
Here $T_i$ is the initial temperature at which our baryogenesis mechanism began to work.  We should note that when the temperature is higher than $T_i$, the content of the universe is described by a higher-scale theory, such as the left-right symmetric model \cite{Li:2020eun}, SUSY models \cite{Csaki:1996ks}, or reheating theory, etc. The SM plus ALP is only an effective theory below $T_i$. 

Figs. \ref{fig:nbstfg} show the final generating baryon asymmetry $n_B/s$ with different parameters. We have set the baryogenesis to stop at temperature $T_f=159.5{\rm GeV}$. In the left plot, we have chose $\beta_1=0.1$ and constrained $Y_{PQ}$ by $\Omega_a=\Omega_{DM}$ Using Eq. (\ref{eqomega}). The red line corresponds to $n_B/s=10^{-10}$. As the left plot shows, in range $10^8{\rm GeV}<f_a<10^{11}{\rm GeV}, 10^5{\rm GeV}<T_i<10^7{\rm GeV}$, there exist parameters to generated the desired $n_B/s$.
In the right plot, we have set $Y_{PQ}=40$ and $f_a=10^9{\rm GeV}$. The coupling parameter $\beta_1$ is related to the axion-photon coupling by $g_{a\gamma\gamma}=\beta_1\cos^2\theta_W/f_a$, where $\theta_W$ is the Weinberg angle. The Blue line indicates the constraint $g_{a\gamma\gamma}<6.6\times 10^{-11}{\rm GeV}$ by CERN Axion Solar Telescope (CAST) \cite{CAST:2017uph}. To obtain the desired $n_B/s$ and not violate the CAST constraint, $T_i$ should be larger than  $3\times 10^5{\rm GeV}$, as the right plot shows. 

In Fig. \ref{fig:plot33}, the excluded regions in ALP parameter space are shaded. The limits are taken from the \texttt{AxionLimits} \cite{AxionLimits}. The CAST gives the constraint $g_{a\gamma\gamma}<6.6\times 10^{-11}{\rm GeV}^{-1}$ when $m<0.02{\rm eV}$. The
study of the evolution of the horizontal branch (HB) stars gives similar bounds but for a larger APL mass \cite{Ayala:2014pea}. APLs with large mass can decay to photons, which is constrained by the galactic X-ray spectra and the extragalactic background light (EBL) \cite{Overduin:2004sz}. The photons from ALP decay can lead to ionization of primordial hydrogen, which constrains the ALP-photon coupling $g_{a\gamma\gamma}$ as the purple region in Fig. \ref{fig:plot33} shows.  The dashed lines in Fig. \ref{fig:plot33} show the   $n_B/s$ for different $g_{a\gamma\gamma}$. We have taken parameters $\Omega_a=\Omega_{\rm DM},T_i=10^6{\rm GeV}$. The final baryon asymmetry $n_B/s$ increases when $g_{a\gamma\gamma}$ increases, as Fig. \ref{fig:plot33} shows. There exist parameters that achieve the observed  $n_B/s$ and are not excluded by the current data.

\begin{figure}[htbp]
    \centering
    \includegraphics[width=0.49\linewidth]{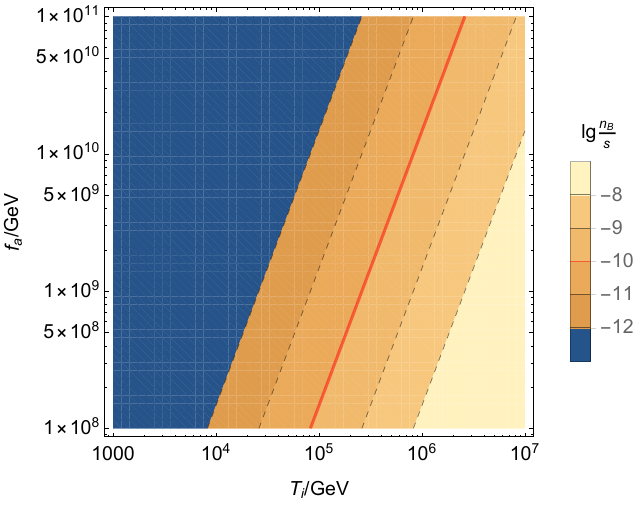}
    \includegraphics[width=0.49\linewidth]{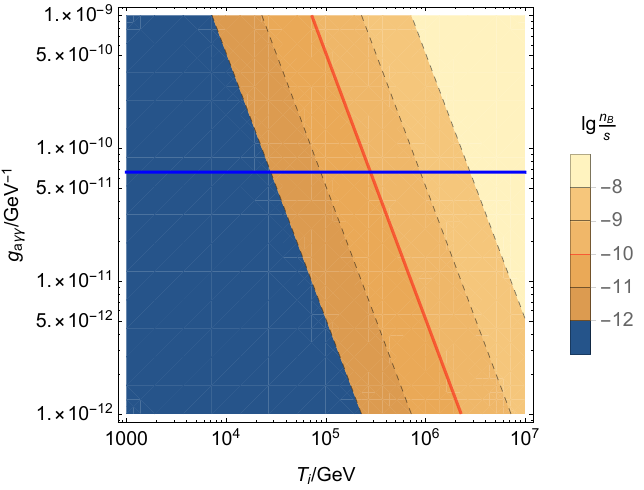}
    \caption{The final generated baryon asymmetry with different parameters. We have taken $T_f=159.5{\rm GeV}$. In the left plot, we have chose $\beta_1=0.1$ and constrained $Y_{PQ}$ by $\Omega_a=\Omega_{DM}$. The red line corresponds to $n_B/s=10^{-10}$. In the right plot, we take $Y_{PQ}= 40$ and $ f_a=10^9{\rm GeV}$. The Blue line indicates the constraint $g_{a\gamma\gamma}<6.6\times 10^{-11}{\rm GeV}$ by CAST \cite{CAST:2017uph}.} \label{fig:nbstfg} 
\end{figure}

\begin{figure}[htbp]
    \centering
    \includegraphics[width=0.8\textwidth]{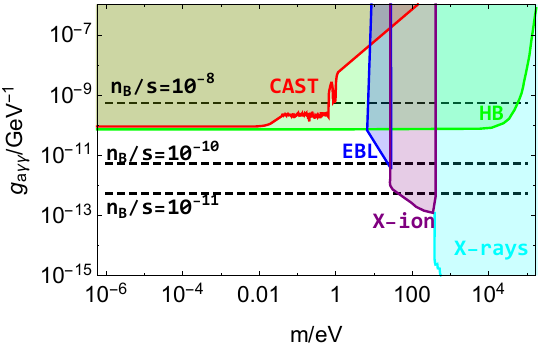}
    \caption{Constraints for $g_{a\gamma\gamma}$. The shaded region is excluded, and these constraints are taken from the repository \texttt{AxionLimits} \cite{AxionLimits}. The dashed lines indicate $n_B/s$ for different $g_{a\gamma\gamma}$ with parameters $\Omega_a=\Omega_{\rm DM},T_i=10^6{\rm GeV},T_f=159.5{\rm GeV}$.  }
    \label{fig:plot33}
\end{figure}

\end{itemize}

\section{ CMB spectral
distortion  from ALPs}\label{sec5}
\begin{figure}[htbp]
    \centering
    \includegraphics[width=0.8\textwidth]{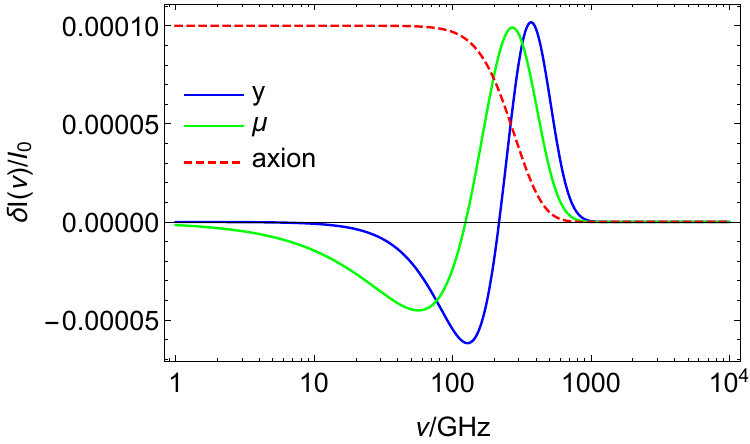}
    \caption{The CMB spectral
distortions. The red dashed line is spectral
distortion from ALPs, while the blue and green lines are $ \ gamma$ distortion and $\mu$ distortion, respectively. We have normalized the  spectral distortions $\delta I$ by $I_0=4\pi(T_0/(2\pi))^3\approx 270{\rm MJy}\, {\rm sr}^{-1}$. We have taken $\mu =9\times 10^{-5},y=1.5\times 10^{-5},\epsilon=10^{-2}$.} 
    \label{fig:distor}
\end{figure}

After the electroweak phase transition, the 
$SU(2)_L\times U(1)_Y$
  symmetry of the SM is spontaneously broken to the $U(1)_{em}$ gauge group,  and the Lagrangian (\ref{couple}) reduce to couplings between ALPs and photons,
\begin{equation}\label{couple2}
 L\supset- g_{a\gamma\gamma}\frac{\phi}{4 } A_{\mu\nu}\tilde{A}^{\mu\nu}.  
 \end{equation}
Similar to Eq. (\ref{ssgx}), the dispersion relation for  photons is 
\begin{equation}\label{ssgx2}
  \begin{aligned}
 \omega_\pm \approx \frac{k}{a}\mp  \frac{g_{a\gamma\gamma}\dot{\phi}}{2}.
  \end{aligned}  
\end{equation}
It should be noted that this relation holds only for frequency $\omega\gg g_{a\gamma\gamma}\dot{\phi}/2$. Since the frequencies of most photons are concentrated around the temperature, this relation holds for the majority of photons as long as 
$g_{a\gamma\gamma}\dot{\phi}\ll T$. 
Assuming that photons are in thermal equilibrium prior to recombination, the photon distribution function is given by
\begin{equation}
 f_\pm(k)=\left(\exp\left(\frac{k}{aT}\mp \frac{g_{a\gamma\gamma}\dot{\phi}}{2T}\right)-1\right)^{-1}.   
\end{equation}
We treat the axion field as a homogeneous background. The number density of photons with momentum $k/a$ can be altered solely through collisions with other particles. Consequently, after recombination, this number density remains conserved, and the photon distribution function is given by
\begin{equation}
 f_\pm(k)=\left(\exp\left(\frac{k}{aT}\mp \frac{g_{a\gamma\gamma}\dot{\phi}_{rec}}{2T_{rec}}\right)-1\right)^{-1},   
\end{equation}
where $T_{rec}\approx 0.26 {\rm eV}$ is temperature at recombination and $\dot{\phi}_{rec}$ means $\dot{\phi}$ when recombination happened. The subscript rec denotes recombination.
The intensity of the photon field observed today is
\begin{equation}
\begin{aligned}
 I(\nu)&=2\pi \nu^3\left(f_+(2\pi \nu)+f_-(2\pi \nu)\right) \\
  &\approx I_b(\nu)+\delta I_a(\nu)\\
\end{aligned}
\end{equation}
where the first term is  the blackbody intensity 
\begin{equation}
 I_b(\nu)= \frac{4\pi \nu^3}{e^{2\pi \nu/T_{0}}-1},  \quad T_{0}=2.726{\rm K}
\end{equation}
and the second term is the spectral distortion
\begin{equation}
 \delta I_a(\nu)=\epsilon^2\frac{2 \pi   \nu^3 e^{2\pi\nu/T_0} \left(e^{2\pi\nu/T_0}+1\right)}{\left(e^{2\pi\nu/T_0}-1\right)^3},\quad \epsilon=\frac{g_{a\gamma\gamma}\dot{\phi}_{rec}}{2T_{rec}}.
\end{equation}
Historically, two types of spectral distortions have been extensively studied: the $\mu$ distortion and the $y$ distortion \cite{Chluba:2025wxp, Lucca:2019rxf, Chluba:2018cww}, which take the forms,
\begin{equation}
 \begin{aligned}
  \delta I_\mu (\nu) &= \mu\cdot 4\pi \nu^3 M\left(\frac{2\pi \nu}{T_0}\right),\quad M(x)=\frac{x e^x}{(e^x-1)^2}\left(0.4561 -\frac{1}{x}\right),\\
  \delta I_y (\nu) &= y\cdot 4\pi \nu^3 Y\left(\frac{2\pi \nu}{T_0}\right),\quad Y(x)=\frac{xe^x }{(e^x-1)^2}\left(x\frac{ e^x+1}{e^x-1}-4\right).
 \end{aligned}   
\end{equation}
The COBE/FIRAS set stringent upper limits $|\mu|<9\times 10^{-5},|y|<1.5\times 10^{-5}$ \cite{Fixsen:1996nj,Kogut:2019vqh}. In Fig. \ref{fig:distor}, the CMB spectral
distortion induced by ALPs is shown by a red dashed line. For comparison, we also show the y distortion and $\mu$  distortion with $\mu =9\times 10^{-5},y=1.5\times 10^{-5}$. All the spectral distortions $\delta I$ are normalized  by $I_0=4\pi(T_0/(2\pi))^3\approx 270{\rm MJy}\, {\rm sr}^{-1}$. We have chosen a sufficiently large $\epsilon=10^{-2}$ to obtain a value of the same order as $\mu$ and y  distortions. The most significant distinction between distortion from axion and the other two types of distortions is that at low frequency, $\delta I_a$ tends to $\epsilon^2 I_0$, while  $\delta I_\mu$ and $\delta I_y$ tend to zero. Notice that $\delta I_a\propto \dot{\phi}_{rec}^2$; the spectral
distortion $\delta I_a$ is always positive even if the axion field is inhomogeneous. 
In the traditional misalignment mechanism, $\dot{\phi}_{rec}^2\sim \rho_{rec}/2=\rho_0 T_{rec}^3/{2T_0^3}$, where $\rho_{rec},\rho_0$ are the energy density of ALPs at recombination and the current value, respectively.  Thus, $\epsilon^2\sim 1.24\times 10^{-5}(g_{a\gamma\gamma}\cdot {\rm GeV})^2({\Omega_a}/{\Omega_{\rm DM}})$, where ${\Omega_a}/{\Omega_{\rm DM}}$ is the proportion of ALPs in dark matter. Notice that for  $g_{a\gamma\gamma}\sim 10^{-10}{\rm GeV}^{-1} $, $\epsilon^2$ is tiny, and the CMB spectral distortion induced by ALPs is difficult to detect.

\section{  conclusions and Discussions}\label{sec4}
In this work, we propose a mechanism to generate baryon asymmetry. We study the $U(1)_Y$ gauge field under a homogeneous ALP background. Due to the interaction with ALP, the two circular polarization modes of the $U(1)_Y$ gauge field have different dispersion relations, which will lead to nonzero $\langle\boldsymbol{E\cdot B}\rangle$. According to the anomaly equation in the SM, the variation of  $\langle\boldsymbol{E\cdot B}\rangle$ can result in the violation of baryon $B$ and lepton $L$  numbers, while the $B-L$ remains unchanged. Assuming the gauge field is in thermal equilibrium, we derive the relation between the baryon number density and the axion background. We estimate the final baryon asymmetry in two scenarios. In the traditional misalignment mechanism, it is difficult to generate enough baryons due to the small $|\dot{\phi}|/(f_a T)$ for axion mass $m<1.1\time 10^{-4}{\rm eV}$ or the inhomogeneity of the axion background for  $m>1.1\time 10^{-4}{\rm eV}$. In the kinetic misalignment mechanism, the final baryon asymmetry depends on the PQ symmetry charge normalized by the entropy density $Y_{PQ}$, the PQ scale $f_a$, the coupling between the axion and gauge field $\beta_1$, and the initial temperature $T_i$ at which our baryogenesis mechanism began to work. We have scanned the parameter space and identified parameters that yield the desired $n_B/s$ and dark matter abundance.

In SM, both the $SU(2)_L$ and $U(1)_Y$ gauge fields can change baryon number through anomaly. In this work, only the $U(1)_Y$ gauge field with axion coupling is considered. If the  $SU(2)_L$ gauge fields and axion have nonzero coupling, the contribution from $SU(2)_L$ gauge fields should be considered, which is beyond the scope of the present work. An interesting result is that, in the kinetic misalignment mechanism, the final baryon asymmetry depends on the initial temperature $T_i$ at which our baryogenesis mechanism begins to operate. A large $T_i$ may lead to excessive baryon numbers, which implies the SM is only an effective theory below $T_i$. In our estimates, $T_i$ can range from $10^5$ to $ 10^7$ GeV; above that temperature, the content of the universe should be described by a higher-scale theory.  

After the electroweak phase transition, the couplings between ALPs and gauge fields of $SU(2)_L\times U(1)_Y$ are reduced to those between ALPs and photons. We show that the axion background can alter the photon dispersion relation, thereby leading to CMB spectral distortion. The spectral distortion $\delta I_a$ is estimated. we found that it approachs to a constant $\epsilon^2 I_0$, which is significantly different with conventional y and $\mu$  distortions. Due to the faint coupling between ALPs and photons, the spectral distortion is small and difficult to detect. A similar discussion applies to any other pseudoscalar coupled to photons that can produce CMB spectral distortions via a similar mechanism,  which may be less constrained and have a bigger distortion effect.

%\begin{acknowledgments
%.....  
%\end{acknowledgments}

\bibliographystyle{unsrt}
\bibliography{myrefs.bib}
\end{document}